\documentclass[
    ,final            
  ]
  {aipproc}

\def\lognlogs{Log~N~--~Log~S\ }
\mathchardef\mhyphen="2D

\layoutstyle{6x9}

\begin{document}

\title{Population synthesis of DA white dwarfs:\\ constraints on soft X-ray spectra evolution}

\author{P.~A. Boldin}{
  address={National Research Nuclear University "MEPhI", Kashirskoe shosse 31, Moscow 115409, Russia},
  altaddress={Stenberng Astronomical Institute, Universitetski pr. 13, Moscow 119991, Russia},
email={boldin.pavel@gmail.com},
}
\author{V.~F. Suleimanov}{
  address={Insitute for Astronomy and Astrophysics, Kepler Center for Astro and Particle Physics, Eberhard Karls University, Sand 1, 72076 T\"ubingen, Germany},
altaddress={Kazan Federal University, Kremlevskaya str. 18, 42008 Kazan, Russia},
}
\author{S.~I. Blinnikov}{
  address={Institute for Theoretical and Experimental Physics (ITEP), Moscow 117218, Russia },
  altaddress={Stenberng Astronomical Institute, Universitetski pr. 13, Moscow 119991, Russia},
}
\author{S.~B. Popov}{
  address={Stenberng Astronomical Institute, Universitetski pr. 13, Moscow 119991, Russia},
}

\begin{abstract}
   Extending the population synthesis method 
   to isolated young cooling white dwarfs we are able to confront our model assumptions
   with observations made in ROSAT All-Sky Survey \cite{flem:96}.
   This allows us to check model parameters such as evolution of spectra and separation of heavy elements in 
   DA WD envelopes.
   It seems like X-ray spectrum temperature of these objects is given by the
   formula \cite{wolff:96} $T_{\rm X \mhyphen ray}~=~\min\left(T_{\rm eff},
   T_{\rm max}\right)$.
 We have obtained DA WD's birth rate and upper limit of the X-ray spectrum
   temperature: DA birth rate $= 0.61\times
   10^{-12}\,{\rm pc^{-3}\,yr^{-1}}$, $T_{\rm max} = 4.1\times 10^4\,{\rm
   K}$.  These values are in good correspondence with values obtained
   by \cite{liebert:04,wolff:96}.
 From this fact we also conclude that our population synthesis method is
   applicable to
 the population of close-by isolated cooling white dwarfs as well as to the
   population of the isolated cooling neutron stars.

\end{abstract}

\maketitle


\section{Population Synthesis of isolated compact objects}

Population synthesis is a powerful tool in astrophysics \cite{pp07}.
For example, it was successfully applied to the population of close-by
cooling isolated neutron stars (see \cite{popov:10} and references therein).  
The basics of the model are the following. After initial distributions and
evolutionary laws for the population under study are specified, a
Monte-Carlo calculations of the evolution are made to produce an artificial
population of sources. Then, some properties of this population are
confronted with the available observational data. Typically, if one (or even
several) ingredients are not well known, then comparison with observations can result in
important constraints of these poorly defined parameters. 

In the case of
close-by cooling isolated neutron stars the ingredients include: initial
spatial distribution, birth rate, mass spectrum, initial velocity
distribution, Galactic potential, cooling history (for a given mass),
interstellar absorption, detector properties. The main observational data
on close-by NSs useful for population synthesis studies (a uniform set of data) is due to
the ROSAT All-Sky Survey. In our previous studies we mailny compared the
observed and simulated \lognlogs distributions, where S is a count rate
for a given detector and N is a number of sources
    with count rate larger than a given value 
 ($N(>S)$). Among the ingredients cooling curves are the most uncertain.
So, the population synthesis can be used as a test for the models of
thermal history of neutron stars \cite{p06}.

As the population synthesis of close-by cooling isolated neutron stars is a
well-established field, \footnote{Interested readers can look at the on-line version of the population
synthesis of isolated close-by cooling NSs:
http://www.astro.uni-jena.de/Net-PSICoNS/ (Boldin, Popov, Tetzlaff, in
press).} we want to expand this approach to study a similar class of
sources -- close-by cooling isolated white dwarfs (WDs).

\section{Ingredients for WD Population Synthesis}


    Here we present results only for DA WDs, as mostly sources of this type are X-ray bright.
    The main ingredients of our populations synthesis model are the following:

    {\bf Spatial distribution and birth rate:} 
	We use two different spatial distributions:
	\begin{enumerate}
	    \item
	 double-exponential disc: $n_0~\sim~\exp\left(-|z|/z_\mathrm{scale}\right)\exp\left(-R/R_\mathrm{scale}\right)$, where
	$R$ and $z$ are cylindrical coordinates with the origin at the Galactic center,
	$z_\mathrm{scale} = 250$~pc, $R_\mathrm{scale} = 3000$~pc;
	    \item
		spatial distribution from \cite{robin:04} (one with $t = 7-10$~Gyr):
	 $\rho_0/d_0 \times \{\exp(-(0.5^2+a^2/h_{R_+}^2)^{1/2})-\exp(-(0.5^2+a^2/h_{R_-}^2)^{1/2})\}$,
	 where $a^2 = R^2+\frac{z^2}{\epsilon^2}$, $R$ and $z$ are cylindrical coordinates with the origin at the Galactic center,
	 $\epsilon$ is ellipsity of the distribution equals to $0.0791$ in our case, $h_{R_+}$ = 2530 pc, $h_{R_-}$ = 1320 pc;
	 \end{enumerate}

	As a starting point for fitting the birth rate of DA WDs we take
	the standard value $10^{-12}\ \mathrm{pc^{-3}\,yr^{-1}}$, DA WDs form
	$\approx$ 60\% of the whole population \cite{liebert:04}.


    {\bf Mass distribution:}
        The mass distribution is very important as the cooling history of a WD depends on its mass.
        A mass also influences the chemical composition of a white dwarf and its spectrum. 
	Birth rates of WDs with different masses are significantly different.
        

	We take the mass distribution from \cite{kepler:06},  $M_\mathrm{mean} = 0.6\ M_\odot$. It gives us a good
	first-order approximation. For a more detailed study we plan to use a spatially dependent mass distribution.

    {\bf Cooling curves:}
	For white dwarfs cooling curves are known much more precisely than
for neutron stars.
        Here we do not put constraints on thermal history of WDs.
	We use cooling curves of WDs computed with the code {\sc Stella} described in \cite{BDB:94}.


    {\bf White dwarfs X-ray spectra:}
	As it appeared, the most uncertain part of our population synthesis
	model is related to X-ray spectra of WDs.
 This is so because even for moderately hot WDs settling of the heavy
	elements is not possible due to radiative levitation,
 and heavy elements contribute more to the opacity in the soft X-ray band. 
	According to our present simple model, 
 the population synthesis shows good correspondence with observations if the
 X-ray spectra temperatures are assumed to follow the formula: 
 $T_\mathrm{X\mhyphen ray} = \min\left(T_\mathrm{eff},
	T_\mathrm{max}\right)$.

	The X-ray spectra used here are computed by the code described in \cite{sul:06}.

	From the final \lognlogs it is clearly seen, that the spectral
        evolution is the 
 most important ingredient of our population synthesis
 study, and confronting the results of calculations with observations one
        can constrain the spectral evolution.

	Because a spectrum of a WD is flatter in soft X-rays than the
	blackbody spectrum, it is important to incorporate a good absorption
	model and the ISM distribution.  Our investigation has shown that
	abundances of elements are also very important, as, e.g.  C and O
	contribute a lot to opacity in the energy range of interest.  We
	used abundances from \cite{wilms:00}.  According to our results,
	change in power-law index is caused by change of number density of
	ISM matter (Local Bubble vs.  near-by Galaxy).

\section{Results}

\begin{figure}
    \includegraphics[height=0.3\textheight]{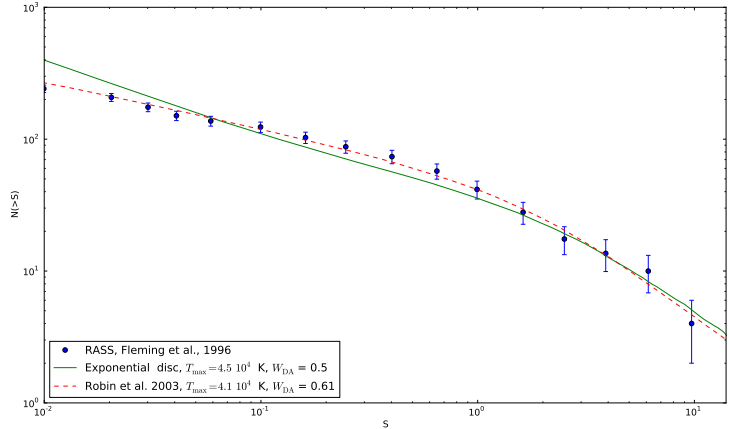}
  \caption{ 
    Our best-fit \lognlogs models for different spatial distributions. Solid line is for the double-exponential disc with
    fitted values $T_\mathrm{max} = 45000$~K and DA WD birth rate $ = 0.5\times 10^{-12}\,{\rm pc^{-3}\,yr^{-1}}$.
    Dashed line is distribution from \cite{robin:04}
    with fitted values $T_\mathrm{max} = 41000$~K and DA WD birth rate $ = 0.61\times 10^{-12}\,{\rm pc^{-3}\,yr^{-1}}$,
    that is overall best-fit model.
}
\end{figure}

\begin{figure}
   \includegraphics[width=0.4\textwidth]{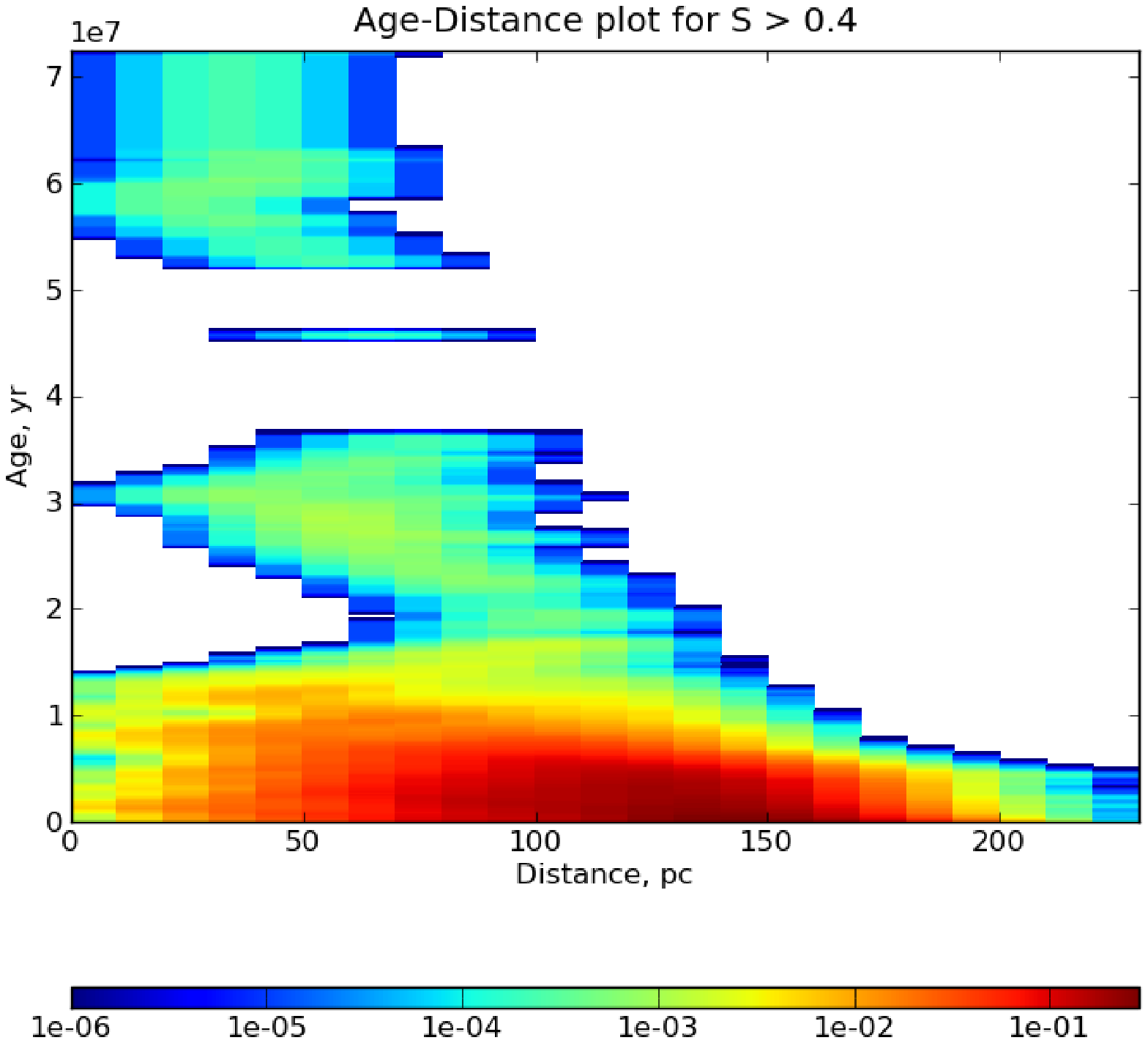}
   \includegraphics[width=0.4\textwidth]{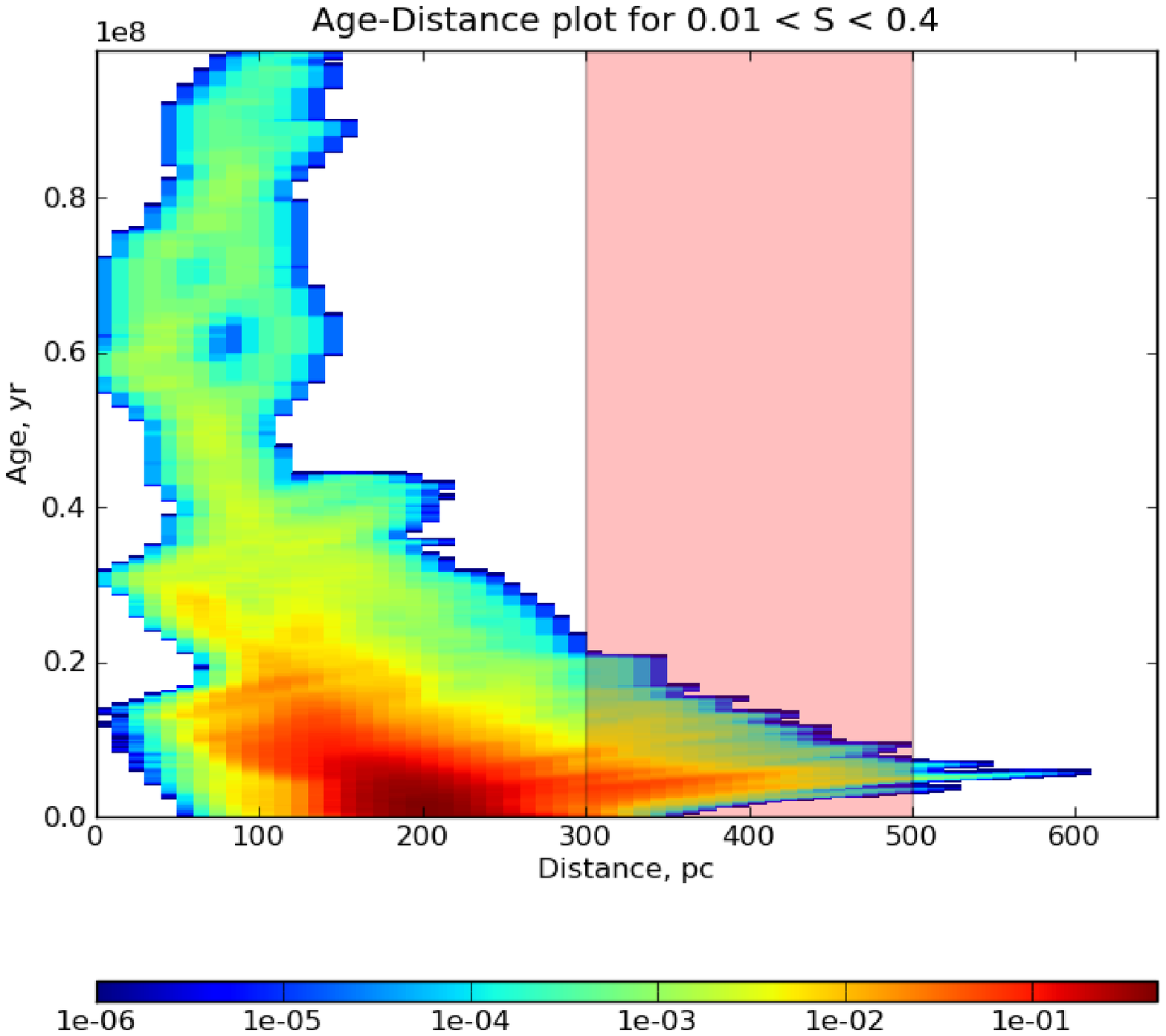}
  \caption{ 
{\it Left:} Age-Distance plot (number density in the 
plane age-distance of objects with given S) for bright sources ($S >
0.4\,\mathrm{cts/s}$).
 {\it Right:} Age-Distance plot for 
intermediate bright objects where the model underpredicts the observed
numbers ($0.01 < S < 0.4$).
 The shaded region corresponds to the Gould Belt.
}
\end{figure}

     In  Fig. 1 we show two best-fit  (in terms of $T_\mathrm{max}$, $W_\mathrm{DA}$) \lognlogs 
     curves for two different spatial distributions. Our best-fit model
     is for the spatial distribution from Robin et al. \cite{robin:04}
     and gives us $T_\mathrm{max} = 41000$~K and $W_\mathrm{DA} = 61\%$.
     These values are in surprisingly good correspondence with values obtained independently (\cite{wolff:96,liebert:04})

      In  Fig. 2 we show two age-distance distributions of calculated WDs.
      Both panels are plotted for the best model with distribution from Robin et al. \cite{robin:04} and $T_\mathrm{max} = 4.1\,10^4\,K$
      (dashed on the \lognlogs plot).
      Shaded areas mark the distance range of the Gould Belt  ($300$~pc~$ < D < 500$~pc).



\begin{theacknowledgments}
 
We thank Dr. Valery Hamrayan for his suggestion to apply the population
synthesis method to cooling WDs and Natalya Dunina-Barkovskaya, who
contributed to the code {\sc Stella}.
PB thanks the organizers for support.
The work of PB and SP is supported by the RFBR and the Federal program for
scientific staff 02.740.11.0575.
The work of VS is supported by the DFG grant SFB / Transregio 7 ``Gravitational Wave Astronomy''
and the RFBR grant  09-02-97013-p-povolzhe-a.
\end{theacknowledgments}

\newcommand{\pasp}{PASP}
\newcommand{\araa}{ARA\&A}
\newcommand{\aplett}{ApL}
\newcommand{\apj}{ApJ}
\newcommand{\apjs}{ApJS}
\let\apjl\apj
\newcommand{\aj}{AJ}
\newcommand{\apss}{Ap\&SS}
\newcommand{\aap}{A\&A}
\newcommand{\aapr}{A\&A. Rev.}
\newcommand{\prd}{Phys.\ Rev.\ D}
\newcommand{\pre}{Phys.\ Rev.\ E}
\newcommand{\physrep}{Phys.\ Rep.}
\newcommand{\iaucirc}{IAU Circ.}
\newcommand{\mnras}{MNRAS}
\newcommand{\nat}{Nature}
\def\pasj{PASJ}

\bibliography{pboldin}
\bibliographystyle{aipproc}






\end{document}